# Ultra-high oxidation potential of Ti/Cu-SnO2 anodes fabricated by spray pyrolysis for wastewater treatment

Aqing Chen [1*], Xudong Zhu [1], Junhua Xi [1], Haiying Qin [1], Zhenguo Ji [1]

[1] *College of Materials & Environmental Engineering, Hangzhou Dianzi University, Hangzhou 310018, P R China.*

**Abstract**

High oxidation potential indicates anodes have high oxidation power and high current efficiency for organics oxidation. In this paper, we prepared Ti/Cu-SnO$_2$ anodes using spray pyrolysis method. The Ti/Cu-SnO$_2$ anodes have an ultra-high oxidation potential of about 2.7 V *vs* NHE, suitable for use in wastewater treatment. The Ti/Cu-SnO$_2$ anodes exhibit the preferred orientation along (110) plane at high Cu doped concentration. First-principles calculations suggest that work function of Cu-SnO$_2$ increases with the Cu doping concentration. It is obtained that the preferred (110) plane and the high work function are responsible for the enhanced oxidation potential of Ti/Cu-SnO$_2$ anodes.

## 1. Introduction

Electrochemical advanced oxidation (EAO) processes which are recognized as the next generation technologies for the wastewater treatment [1] due to the high oxidation efficiency, fast reaction rate and easy operation. The oxidation power of the anodes has great impacts on the effective of the EAO processes. For the low oxidation power anode such as IrO$_2$-based electrodes[2,3], the interaction between electrode and hydroxyl radical is strong, which results in a low current efficiency for organics oxidation. Compared to the low oxidation power anodes, the high oxidation power anodes including $SnO_2$-based electrodes [4–8] and boron-doped diamond (BDD) [9,10] have weaker interaction with the hydroxyl radical leading to high current efficiency for organics oxidation. The oxidation power of the anodes is determined by oxidation potential corresponding to the onset potential of oxygen evolution.

---

[*] Corresponding author.
E-mail address: aqchen@hdu.edu.cn

Titanium-based $SnO_2$ anodes as a kind of mixed metal oxide (MMO) electrodes have gained growing attention due to the relative high oxidation potential. Intrinsic $SnO_2$ is an *n*-type semiconductor with poor conductivity. Used as the anodes, the conductivity of $SnO_2$ can be improved significantly by doping Sb or F [11–13]. It is well known that the Ti/Sb-$SnO_2$ anodes have the oxidation potential from 1.9 to 2.2 V [14–17] which is higher than that of other MMO electrodes. However, at present, the oxidation potential of $SnO_2$-based anodes is still lower than that of BDD [18], although Zn/Sb co-doping can improve the oxidation potential of $SnO_2$-based anodes up to 2.4 V [19].

In this present work, we demonstrate titanium-based Cu doped $SnO_2$ (Ti/Cu-$SnO_2$) anodes exhibiting a ultra-high oxidation potential of about 2.7 V (*vs* NHE) which is comparable to that of BDD (2.7 V *vs* NHE) [10]. The spray pyrolysis method as the general technique for preparing doped $SnO_2$ thin films [20,21] was used to fabricate the Cu doped $SnO_2$ anodes. Moreover, the relationship between oxidation potential and the Cu doping concentration was discussed and detailed reasons for the ultra-high oxidation potential were explored further by x-ray diffraction (XRD) analyses and density functional theory (DFT) calculations.

## 2. Experimental details

The Ti/Cu-$SnO_2$ anodes were prepared by the deposition of the Cu doped $SnO_2$ layer on Ti substrates of 2 cm × 2 cm using a homemade spray pyrolysis apparatus. The Ti substrates were pretreated by sandblasting and then etched in boiling 10% oxalic acid during 30 min. The precursor solutions for Cu doped $SnO_2$ layer were prepared by adding $SnCl_4$-$5H_2O$ and $Cu(NO_3)_2$ into ethanol. The amount of $SnCl_4$-$5H_2O$ is 8.7 g for all samples. The Cu concentration in precursor solution is 1.1 at.%, 4.2 at.% and 8.1 at.% corresponding to sample 1 (S1), sample 2 (S2) and sample 3 (S3), respectively. The resultant solutions were sprayed on the pretreated Ti substrates at the 550 ± 5 °C temperature at which the $SnO_2$-based anodes have high durability [8,15]. The height between the spray nozzle and Ti substrates (3 cm) and the flow of carrier gas (100L/h) were kept fixed. For comparison, Ti/Sb-$SnO_2$ electrodes have been prepared with the precursor solutions containing $SnCl_4$-$5H_2O$, $Sb_2O_3$ and several drops of HCl. The Sn:Sb atomic ratio is 9.0:1 in the precursor solution.

The crystal structure of the Cu doped $SnO_2$ thin films was characeriszed using X-ray diffraction (XRD) diffractometer with Cu $K_\alpha$ radiation, with a scanning angle (2θ) range of 20° to 50°. The scanning electron microscopy (HITACHI S4800) was employed to examine the morphology of Ti/Cu-$SnO_2$ anodes with different Cu doping level. The cyclic voltammetry (CV) tests using a standard three electrode cell were carried out to measure the oxidation potential of Ti/Cu-$SnO_2$ and Ti/Sb-$SnO_2$ anodes. The electrolyte was 0.5 M $H_2SO_4$ solution. Pt plate with the area of 1×1 $cm^2$ was used as a counter electrode and Hg/$Hg_2SO_4$.$K_2SO_4$ (0.64 V *vs* NHE) as a reference electrode. The working electrodes were the Ti/Cu-$SnO_2$ and Ti/Sb-$SnO_2$ anodes.

## 3. Results and discussion

We quantify the Cu content in the anodes by energy dispersive x-ray spectroscopy (EDS). The Cu concentrations in the anodes prepared by the precursor solution with Cu content of 1.1 at.%, 4.2 at.% and 8.1 at.% are 0.5 wt.%, 1.1 wt.% and 1.9 wt.%, respectively. We measured the carrier concentrations of the Cu doped $SnO_2$ by a Hall effect measurement system at room temperature, as shown in Fig. 1. The carrier density in the Cu-$SnO_2$ decreases with Cu doping level due to the compensation of the intrinsic donors by the Cu acceptors [22]. Besides, Hall effect measurements also reveal that the conduction of Cu doped $SnO_2$ thin films in the present work is still *n* type.

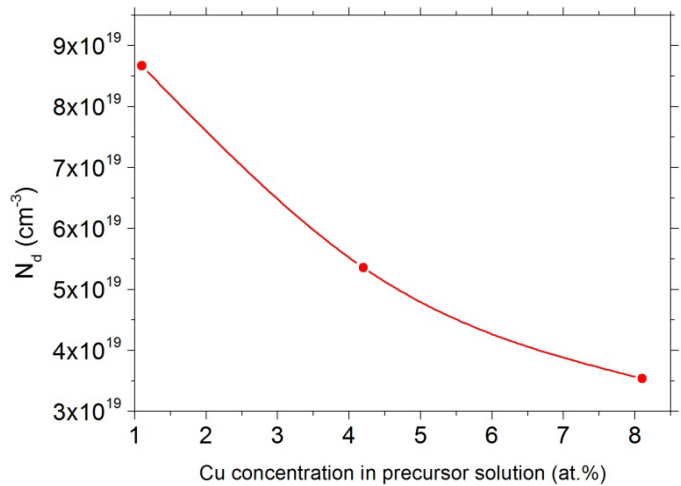

Fig. 1 the carrier density as a function of the Cu concentraion in precursor solution

Fig. 2a, 2b and 2c present the scanning electron microscopy (SEM) images of Cu-SnO$_2$ thin films prepared by spray pyrolysis at 550 ± 5 °C with the Cu doping concentration of 1.1 at.%, 4.2 at.% and 8.1 at.% in spray solution, respectively. No cracks on the surface of Cu-SnO$_2$ thin films are observed. It can be seen from Fig. 2a and 2b that the Cu-SnO$_2$ thin films are compact with an average particle size of ~ 0.5 μm. But the average particle size of Cu-SnO$_2$ thin films increases up to ~1 μm for the Cu doping concentration of 8.1 at.% in precusor solution. Moreover, it novel to observe that the grain become diamond-shaped, which is quiet different from the SnO$_2$ thin films with the Cu doping concentration of 1.1 at.% and 4.2 at.% in precursor solution. This special surface morphology can affect the oxidation potential of Ti/Cu-SnO$_2$ anodes, which has been proven by the the cyclic voltammograms measurements below.

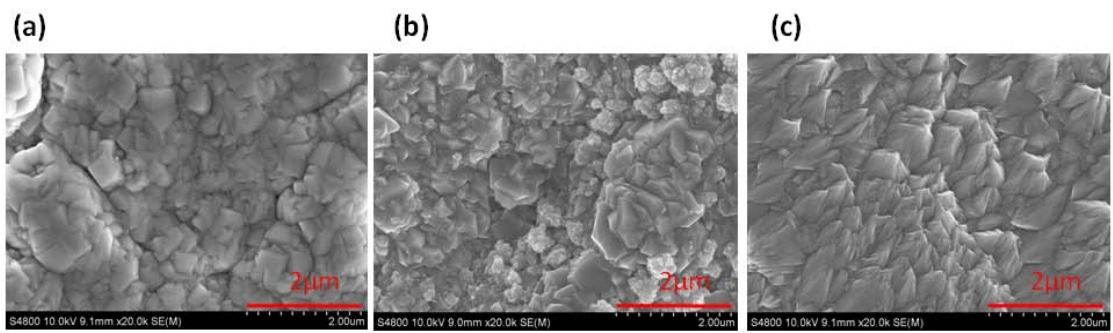

Fig. 2 SEM images of Ti/Cu-SnO$_2$ electrodes (×20000) with the Cu doping concentration of (a) 1.1 at.%, (b) 4.2 at.% and (c) 8.1 at.% in the precursor solution.

Fig. 3 shows the cyclic voltammograms (CV) of Ti/Sb-SnO$_2$ and Ti/Cu-SnO$_2$ electrodes in 0.5 M H$_2$SO$_4$ solution at a scan rate of 100 mV/s. It can be obtained the oxidation potential of Ti/Sb-SnO$_2$ is about 2.2 V *vs* NHE, which agrees well with reported value [23]. It is novel to find that the oxidation potential of Ti/Cu-SnO$_2$ is as high as 2.7 V *vs* NHE, which is 0.5 V bigger than that of Ti/Sb-SnO$_2$. It is believed that the interaction between electrode surface and hydroxyl radical determines the oxidation potential [17]. This interaction is related to the surface potential. High surface potential results in a weak interaction between electrode surface and hydroxyl radical. So, the enhanced oxidation potential suggests that the Ti/Cu-SnO$_2$ electrodes have high surface potential.

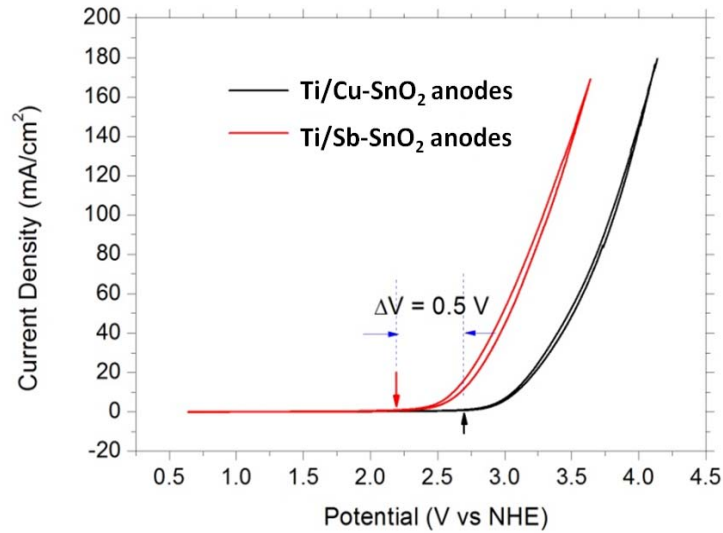

Fig. 3 Polarization curves for oxygen evolution obtained on the Ti/Cu-SnO$_2$ and Ti/Sb-SnO$_2$ electrodes in 0.5 M H$_2$SO$_4$ solution at a scan rate of 100 mV/s. For Ti/Sb-SnO$_2$ electrodes the Sb concentration is 10 at. % in precursor solution and for Ti/Cu-SnO$_2$ electrodes the Cu concentration is 8.1 at. % in precursor solution.

Fig. 4 shows the cyclic voltammograms of Ti/Cu-SnO$_2$ electrodes with different Cu doping concentrations. It is worth noting that the oxidation potential of S1 (V$_{OP1}$) is about 2.2 V *vs* NHE which is in well agreement with that of Ti/Sb-SnO$_2$ electrodes. Increasing the Cu doping concentration up to 4.2 at.% leads to the oxidation potential of S2 (V$_{OP2}$) of about 2.6 V *vs* NHE. As the Cu doping concentration reaches 8.1 at.%, a ultra-high oxidation potential of 2.7 V *vs* NHE is obtained. So, it is obtained that the oxidation potential increases with the Cu doping concentration.

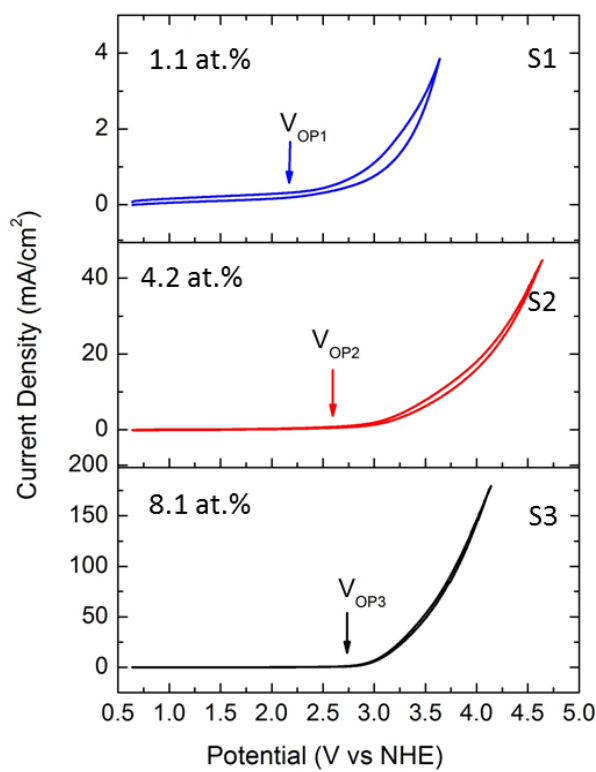

Fig. 4 Polarization curves for oxygen evolution in 0.5 M H$_2$SO$_4$ solution at a scan rate of 100 mV/s for the Ti/Cu-SnO$_2$ electrodes with different Cu concentrations in precursor solution (at.%). V$_{OP1}$, V$_{OP2}$ and V$_{OP3}$ are the oxidation potential of Ti/Cu-SnO$_2$ anodes with Cu concentrations of 1.1 at.% (S1), 4.2 at.% (S2) and 8.1 at.% (S3) in precursor solution, respectively.

As discussed above the Cu doped SnO$_2$ have high surface potential which is affacted by surface crystal structure. In order to obtain the surface crystal structure of Cu doped SnO$_2$ thin films, the XRD measurements were performed on the Cu doped SnO$_2$ with the Cu concentration of 1.1 at. %, 4.2 at. % and 8.1 at. % in precursor solution. The XRD patterns fitted by Gaussians, as shown in Fig. 5, indicate that all of the Cu doped SnO$_2$ films have the rutile structure in a good agreement with JCPD 01-070-4176 and that no impurity phase was observed. Nevertheless, Cu doping has great impacts on the crystal structure of SnO$_2$. As can be seen in Fig. 5, for Ti/Cu-SnO$_2$ anodes with Cu concentration of 1.1 at.% and 4.2 at.% three main characteristics diffraction peaks are at 26.58º, 33.87º and 37.94º corresponding to (110), (101) and (200) planes of rutile structure of the SnO$_2$ phase. Moreover, it is interesting to note that the (101) peak is disappeared for Ti/Cu-SnO$_2$ anodes with the Cu concentrations of 8.1 at.% in precursor solution, which may result from the fact that the Cu doping changes the phase formation or cause local disorder at high concentration of 8.1 at. % in precusor solution.

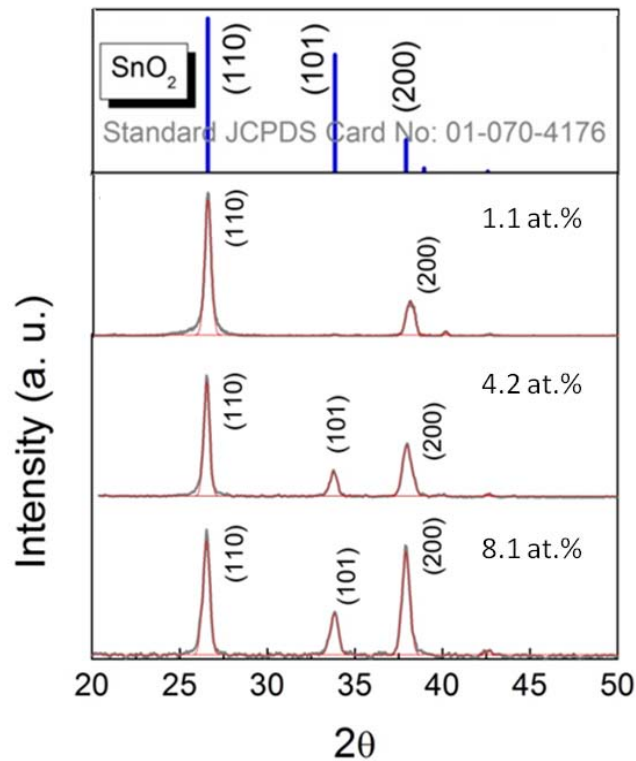

Fig .5 The XRD patterns of the Cu doped SnO$_2$ with the Cu concentration of 1.1 at. %, 4.2 at. % and 8.1 at. % in precursor solution in the 2θ range of 20º to 54º

In order to understand the effects of Cu doping on the crystal orientation, the texture coefficient (TC) factor [24] was calculated by the following equation:

$$TC_{(hkl)} = \frac{I_{(hkl)}/I_{0(hkl)}}{\left(\frac{1}{N}\right)\sum_{N=1}^{N} I_{hkl}/I_{0(hkl)}}$$

where $I_{hkl}$ is the measured intensity value of the *hkl* plane and $I_{0(hkl)}$ is the standard intensity values of the *hkl* plane. N is the number of obtained diffraction peaks in the XRD profile. High $TC_{hkl}$ suggests the preferred growth. Fig. 6 shows the $TC_{hkl}$ of S1, S2, S3 and the standard $SnO_2$ powder (JCPD 01-070-4176). It is obvious to see that the $TC_{110}$ increases with the Cu concentration, and on the contrary, the $TC_{200}$ decreases with the Cu concentration. Both S1 and S2 have a preferred orientation along (200). But it novel to find that S3 exhibits a preferred orientation along (110). The surface potential is related to the crystal strucutre. Therefore, it is obtained that Cu doping can change the surface potential of the $SnO_2$, which ultimately impacts the oxidation potential of Ti/Cu-$SnO_2$ electrodes.

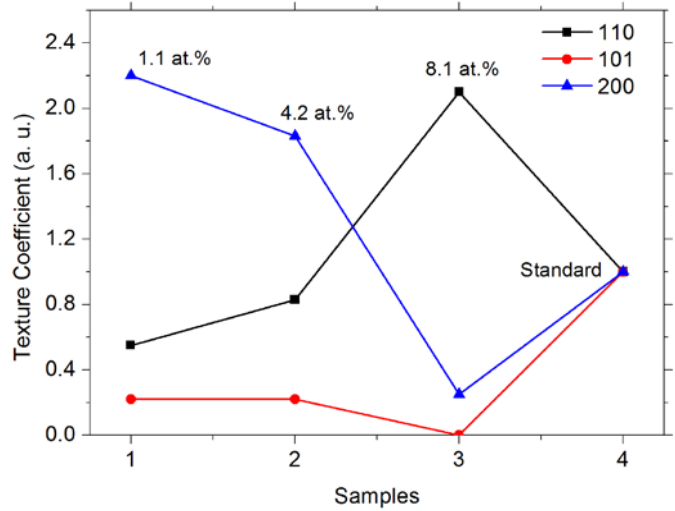

Fig .6 Variation in TC of the Cu-$SnO_2$ with the Cu concentration of 1.1 at. % (Sample1), 4.2 at.% (Sample2), 8.1 at.% (Sample3) in precursor solution and the standard powder (JCPD 01-070-4176, sample 4).

In order to better understand the influences of Cu doping on the oxidation potential, we performed first-principles calculations on the surface potential of $SnO_2$ using the Quantum ESPRESSO package [25]. The exchange-correlation energy of interacting electrons was treated by using the Perdew-Burke-Ernzerhof generalized gradient approximation [26]. The energy cutoff for the plane-wave basis set was 40 Ry. All the models were calculated with a Mokhorst-Pack k-point (4x5x1). As shown in Fig. 7, three type structures with 7 layers separated by 20 Å vacuum layer were chosen to model the $SnO_2$ surface which was constructed by cleaving the $SnO_2$ bulk along (110) plane. They are intrinsic (Fig.7a), one Cu atom doped (Fig. 7b) and 3 Cu atoms doped (Fig. 7c) $SnO_2$.

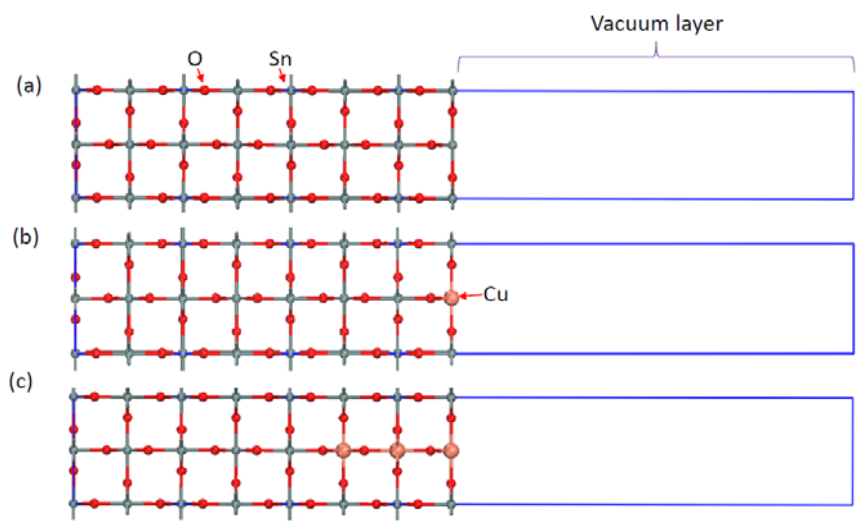

Fig. 7 the calculated models of intrinsic (a), 1 Cu atom doped (b) and 3 Cu atoms doped (c) $SnO_2$ were constructed by cleaving the $SnO_2$ bulk along (110) plane with the thickness of vacuum layer is 20 Å.

The Fermi energy level and work function Φ are the main influence factors of the surface potential. The work function which is the minimum energy needed to remove an electron from the bulk of a material through the surface to vacuum are written by

$$\phi = V_{vac} - E_F \qquad (2)$$

where $E_F$ is the Fermi level and $V_{vac}$ is the vacuum level. Fig. 8a, 8b and 8c shows electrostatic potential energy for intrinsic, 1 Cu atom doped and 3 Cu atoms doped $SnO_2$, respectively. From the plot of electrostatic potential energy, we can obtain the work function of 4.46, 4.87 and 4.93 eV for intrinsic, 1 Cu atom doped and 3 Cu atoms doped $SnO_2$ along (110) plane, respectively. It is obvious to find that the work function increases with the Cu doping concentration. Moreover, the Fermi energy level decreases with the Cu doping concentration, as shown in Fig. 8d. The calculated results on Fermi energy level and work functions indicate that the surface potential of $SnO_2$ along (110) plane increases with the Cu doping concentration.

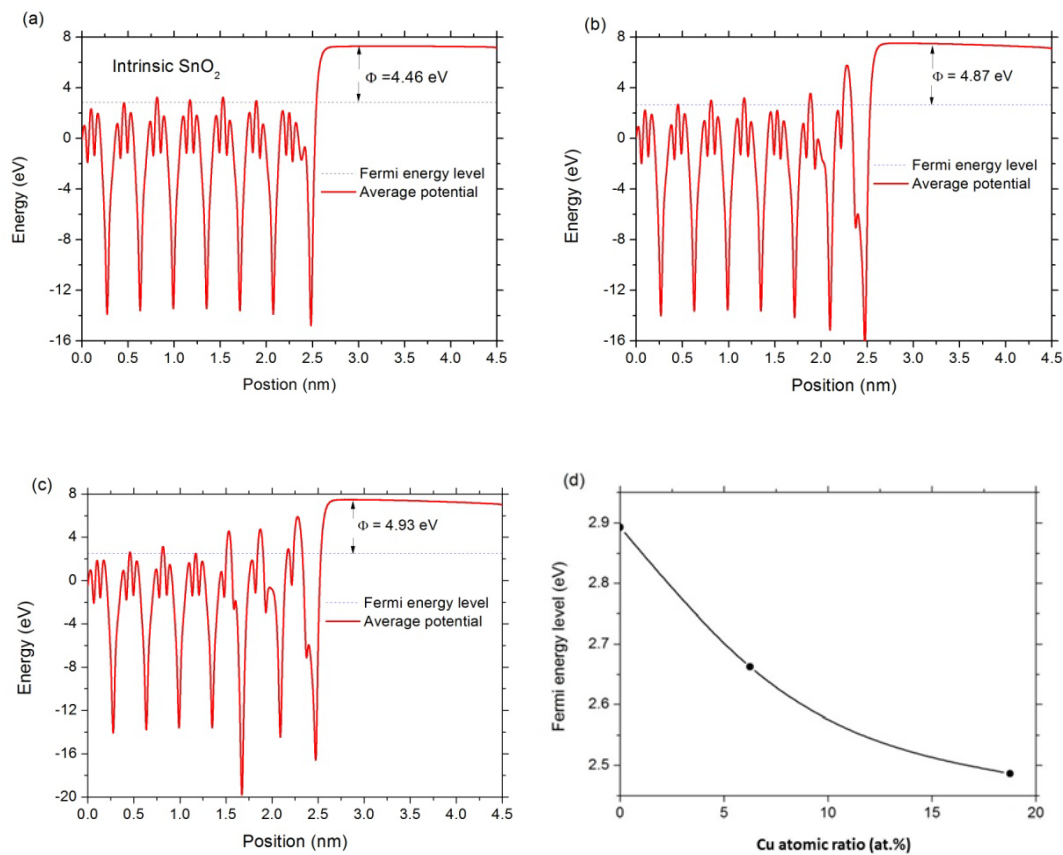

Fig.8 (a) Plane averaged electrostatic potential energy of intrinsic (a), 1 Cu atom doped (b) and 2 Cu atoms doped (c) $SnO_2$ surface along (110) plane, (d) Fermi energy for Cu doped $SnO_2$ as a function of the Cu atomic ratio.

**Conclusion**

In summary, we have prepared the Ti/Cu-$SnO_2$ anodes by spray pyrolysis method. It is found that the Ti/Cu-$SnO_2$ anode has an ultra-high oxidation potential of about 2.7 V *vs* NHE which is comparable to that of BDD. XRD data show that the Cu doped $SnO_2$ thin films trend to show preferred orientation along (110) as the Cu doping concentration increases. The calculated results indicate that the Fermi energy level decreases with Cu doping concentration and that the work function increases with Cu concentration, which further gives the detailed explanations on the enhancement of oxidation potential of Ti/Cu-SnO$_2$ anode. These results have great significance for the development of effective SnO$_2$-based electrodes for wastewater treatment.

http://www.ncbi.nlm.nih.gov/pubmed/10062328.